\documentclass[twocolumn,journal]{IEEEtran}
\usepackage[T1]{fontenc}
\usepackage{bm}
\usepackage{amsmath}
\usepackage{amsthm}
\usepackage{graphicx}
\usepackage[unicode=true,
 bookmarks=true,bookmarksnumbered=true,bookmarksopen=true,bookmarksopenlevel=1,
 breaklinks=false,pdfborder={0 0 0},backref=false,colorlinks=false]
 {hyperref}
\hypersetup{pdftitle={Your Title},
 pdfauthor={Your Name},
 pdfpagelayout=OneColumn, pdfnewwindow=true, pdfstartview=XYZ, plainpages=false}
\usepackage{breakurl}

\makeatletter
\theoremstyle{plain}
\newtheorem{thm}{\protect\theoremname}

\usepackage[caption=false,font=footnotesize]{subfig}

\@ifundefined{showcaptionsetup}{}{%
 \PassOptionsToPackage{caption=false}{subfig}}
\usepackage{subfig}
\makeatother

\providecommand{\theoremname}{Theorem}

\begin{document}

\title{Smart Grid Monitoring Using Power Line Modems: Effect of Anomalies
on Signal Propagation}

\author{Federico~Passerini,~\IEEEmembership{Student Member,~IEEE,} and
Andrea~M.~Tonello,~\IEEEmembership{Senior Member,~IEEE}\thanks{Federico Passerini and Andrea M. Tonello are with the Embedded Communication
Systems Group, University of Klagenfurt, Klagenfurt, Austria, e-mail:
\{federico.passerini, andrea.tonello\}@aau.at.}}
\maketitle
\begin{abstract}
The aim of the present work is to provide the theoretical fundamentals
needed to monitor power grids using high frequency sensors. In our
context, network monitoring refers to the harvesting of different
kinds of information: topology of the grid, load changes, presence
of faults and cable degradation. We rely on transmission line theory
to carry out a thorough analysis of how high frequency signals, such
those produced by power line modems, propagate through multi-conductor
power networks. We also consider the presence of electrical anomalies
on the network and analyze how they affect the signal propagation.
In this context, we propose two models that rely on reflectometric
and end-to-end measurements to extrapolate information about possible
anomalies. A thorough discussion is carried out to explain the properties
of each model and measurement method, in order to enable the development
of appropriate anomaly detection and location algorithms.\end{abstract}


\IEEEpeerreviewmaketitle{}

\section{Introduction}

\IEEEPARstart{H}{igh} frequency monitoring is an essential tool to
operate modern distribution grids, since it allows the utilities to
sense different kind of electrical events that will or potentially
can alter the status of the network. While monitoring is traditionally
performed using phasor measurement units (PMUs) \cite{7961200}, other
devices that can operate at high frequency are nowadays more and more
deployed in distribution grids, e.g. power line modems (PLMs). 

PLMs are conventionally used as communication devices in smart grids
(SGs) \cite{5768099,7467440} but, as shown in this paper, can also
serve as network monitoring devices. This role is similar to that
of DSL Access Multiplexers (DSLAMs) in local loops, for which considerable
amount of research has been carried out (see \cite{1007375,neusphd,333334d5-7017-4468-bab7-59871b51312a}
and references therein). However, the power line medium is rather
different from the twisted pair loops used in DSL, especially when
considering distribution grids. In fact, while DSL loop cables are
standardized for high speed communications, power line cables are
not; the topological structure of power line networks changes rather
often, while DSL networks have constant topologies; the loads of distribution
networks have convoluted frequency profiles and are time variant,
while DSL loads are constant in time and have values close to the
line impedance. Most importantly, the focus in DSL is about providing
high speed internet to the users, so the main interest of the DSL
engineers is to qualify the loop and to detect bad splices or possible
short circuits. The main interest in power distribution networks is,
conversely, to deliver energy in an optimized and reliable way. This
means that it is not only important to identify poor connections and
short circuits, but to prevent them, in order to avoid power cuts
that might cause severe problems to the local community. Wake-up calls
in this context are generated by high impedance faults (HIF) or by
the detection of cable aging due to water treeing, oxidation and other
causes. These events that we name anomalies cause almost undetectable
damage to the network, but on a medium to long run can cause a complete
system failure. Different techniques have been proposed to detect
and locate anomalies, which involve measurements at the mains frequency
and its harmonics up to few kHz, using either pulsed, sinusoidal or
wavelet test signals \cite{faultreview,7468545}.

In recent years, relevant research has been carried out to analyze
what information can be harvested about a power line network (PLN)
using signals from 3 kHz up to 86 MHz, which are typical of power
line communications (see \cite{sgc2017,6295693,6497543} and references
therein). These works rely on the signal generation and acquisition
capabilities of PLMs to sense the grid. Since the research in this
area is still in early stage, there is a need to establish a solid
theoretical foundation about the information carried by high frequency
signals in PLNs.

This paper is dedicated to provide a detailed answer to this requirement.
To this aim, we rely on transmission line theory \cite{Paul:2007:AMT:1554645}
and derive closed-form relations that describe the overall effect
of the network on three different quantities: the impedance and the
reflection coefficient measured at one node (reflectometric sensing),
and the channel transfer function (CTF) of a signal coming from a
far-end node (end-to-end sensing). The differences between these relations
allow us to highlight the information about the network status carried
by each of the aforementioned quantities. Afterwards, we introduce
the presence of electrical anomalies and analytically study how they
affect the propagation of high-frequency signals. In this context,
we present two models that allow us to separate the information relative
to the anomaly from the information relative to the rest of the network.
We discuss the differences between these two models and also analyze
which information about the anomaly can each of the considered quantities
provide. 

The analysis presented in this paper can be used, as done in \cite{SGSII},
to propose efficient modem architectures and post-processing algorithms
to detect, classify and locate anomalies.

The rest of the paper is organized as follows. In Section \ref{sec:Propagation-of-signals}
we provide new insights about the theory that describes the propagation
of signals in power line networks. In Section \ref{sec:Effect-of-anomalies}
we analyze how the presence of an anomaly alters the propagation of
both end-to-end and reflectometric signals. Conclusions follow in
Section \ref{sec:Conclusions}.

\section{Propagation of high frequency signals in PLNs\label{sec:Propagation-of-signals}}

The propagation of high frequency signals in PLNs is generally described
by the multiconductor transmission line (MTL) theory \cite{Paul:2007:AMT:1554645}.
In this section, we introduce some new theoretical insights of this
theory that are needed to relate the input admittance, the input reflection
coefficient and the CTF to the network parameters. 
\begin{figure}[tb]
\begin{centering}
\includegraphics[width=0.9\columnwidth]{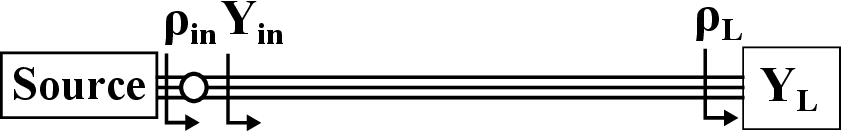}
\par\end{centering}

\caption{Example of a single MTL connecting a signal generator and a load.
\label{fig:Example-of-a-1}}

\end{figure}

Let us consider a single MTL that links a signal generator, which
can be a PLM, and a load described by its admittance matrix $\mathbf{Y_{L}}$
of size $L$x$L$, where $L$ is the number of conductors (see Fig.
\ref{fig:Example-of-a-1}). The propagation of the signal is described
by the telegraph equations
\begin{gather}
\frac{\partial\mathbf{V}\left(x,f\right)}{\partial x}=-\left(\mathbf{R}+j2\pi f\mathbf{L}\right)\mathbf{I}\left(x,f\right)\nonumber \\
\frac{\partial\mathbf{I}\left(x,f\right)}{\partial x}=-\left(\mathbf{G}+j2\pi f\mathbf{C}\right)\mathbf{V}\left(x,f\right)\label{eq:tel}
\end{gather}
where $\mathbf{V}$ and $\mathbf{I}$ are the voltage and current
vectors respectively, $x$ is the distance from the load, $\left\{ \mathbf{R},\mathbf{L},\mathbf{G},\mathbf{C}\right\} $
is the set of matrices that describes the per unit length parameters
of the cable and $f$ is the considered frequency. The dependency
on the frequency will be omitted in the following to simplify the
notation.

\subsection{Analysis of the reflectometry\label{sub:Reflectometry}}

The input admittance matrix $\mathbf{Y}_{\mathbf{in}}$ at the beginning
of an MTL with length $\ell_{1}$ can be written after some derivations
as \cite{versolatto2011an} 
\begin{multline}
\mathbf{Y}_{\mathbf{in}}=\mathbf{T}\left(\mathbf{I}+\mathbf{e}^{-\mathbf{\Gamma}\ell_{1}}\bm{\rho}_{\mathbf{L}}^{\mathbf{M}}\mathbf{e}^{-\mathbf{\Gamma}\ell_{1}}\right)\\
\left(\mathbf{I}-\mathbf{e}^{-\mathbf{\Gamma}\ell_{1}}\bm{\rho}_{\mathbf{L}}^{\mathbf{M}}\mathbf{e}^{-\mathbf{\Gamma}\ell_{1}}\right)^{-1}\mathbf{T}^{-1}\mathbf{Y}_{\mathbf{C}}\label{eq:yinsingle}
\end{multline}
where $\mathbf{Y}_{\mathbf{C}}$ and $\mathbf{\Gamma}$ are the characteristic
admittance and the propagation constant of the cable respectively,
$\mathbf{T}$ is a transformation matrix, $\mathbf{I}$ is the identity
matrix, and $\bm{\rho}_{\mathbf{L}}$ is the load reflection coefficient
matrix defined as 
\begin{equation}
\bm{\rho}_{\mathbf{L}}=\mathbf{Y}_{\mathbf{C}}\left(\mathbf{Y}_{\mathbf{L}}+\mathbf{Y}_{\mathbf{C}}\right)^{-1}\left(\mathbf{Y}_{\mathbf{L}}-\mathbf{Y}_{\mathbf{C}}\right)\mathbf{Y}_{\mathbf{C}}^{-1}\label{eq:plsingle}
\end{equation}
 respectively. We also name $\bm{\rho}_{\mathbf{L}}^{\mathbf{M}}$
the modal load reflection coefficient, where $\bm{\rho}_{\mathbf{L}}^{\mathbf{M}}=\mathbf{T}^{-1}\bm{\rho}_{\mathbf{L}}\mathbf{T}$.
In general, the relation between a modal matrix $\mathbf{A}^{\mathbf{M}}$
and its non-modal counterpart $\mathbf{A}$ is
\[
\mathbf{A}^{\mathbf{M}}=\mathbf{T}^{-1}\mathbf{A}\mathbf{T}.
\]
The input reflection coefficient matrix $\bm{\rho}_{\mathbf{in}}$
is defined as \cite{Paul:2007:AMT:1554645}
\begin{equation}
\bm{\rho}_{\mathbf{in}}=\mathbf{Y}_{\mathbf{R}}\left(\mathbf{Y}_{\mathbf{in}}+\mathbf{Y}_{\mathbf{R}}\right)^{-1}\left(\mathbf{Y}_{\mathbf{in}}-\mathbf{Y}_{\mathbf{R}}\right)\mathbf{Y}_{\mathbf{R}}^{-1}\label{eq:pinsingleeasy}
\end{equation}
where $\mathbf{Y}_{\mathbf{R}}$ is the generator admittance matrix
at the signal generator side. Combining \eqref{eq:yinsingle} and
\eqref{eq:pinsingleeasy}, we obtain 
\begin{equation}
\bm{\rho}_{\mathbf{in}}=\mathbf{N}\mathbf{T}\left(\bm{\rho}_{\mathbf{G}}^{\mathbf{M}}+\bm{\rho}_{\mathbf{B}}^{\mathbf{M}}\right)\left(\mathbf{I}+\bm{\rho}_{\mathbf{G}}^{\mathbf{M}}\bm{\rho}_{\mathbf{B}}^{\mathbf{M}}\right)^{-1}\mathbf{T}^{-1}\mathbf{N}^{-1}\label{eq:pinsingle}
\end{equation}
where $\mathbf{N}=\left(\mathbf{Y}_{\mathbf{R}}+\mathbf{Y}_{\mathbf{C}}\right)\mathbf{Y}_{\mathbf{C}}^{-1}$
and $\bm{\rho}_{\mathbf{G}}^{\mathbf{M}}$ is the modal line mismatch
coefficient matrix defined as
\begin{equation}
\bm{\rho}_{\mathbf{G}}^{\mathbf{M}}=\mathbf{T}^{-1}\mathbf{Y}_{\mathbf{C}}\left(\mathbf{Y}_{\mathbf{C}}+\mathbf{Y}_{\mathbf{R}}\right)^{-1}\left(\mathbf{Y}_{\mathbf{C}}-\mathbf{Y}_{\mathbf{R}}\right)\mathbf{Y}_{\mathbf{C}}^{-1}\mathbf{T}
\end{equation}
and $\bm{\rho}_{\mathbf{B}}^{\mathbf{M}}$ is the modal reflection
coefficient matrix computed at the signal generator side of the line,
defined as
\begin{equation}
\bm{\rho}_{\mathbf{B}}^{\mathbf{M}}=\mathbf{e}^{-\mathbf{\Gamma}\ell_{1}}\bm{\rho}_{\mathbf{L}}^{\mathbf{M}}\mathbf{e}^{-\mathbf{\Gamma}\ell_{1}}.
\end{equation}
We remark that \eqref{eq:yinsingle} and \eqref{eq:pinsingle} are
a multidimensional extension of the scalar input admittance and input
reflection coefficient equations \cite[Ch. 2]{neusphd}, of which
they preserve the structure. Finally, the echo signal coming back
to the signal source from the MTL is given by
\begin{equation}
\mathbf{V_{echo}}=-\mathbf{Y}_{\mathbf{R}}^{-1}\bm{\rho}_{\mathbf{in}}\mathbf{Y}_{\mathbf{R}}\mathbf{V_{source}}.
\end{equation}

In order to derive some useful insights about \eqref{eq:yinsingle}
and \eqref{eq:pinsingle}, we make use of the fact that the Taylor
series of any invertible square matrix $\left(\mathbf{I}+\mathbf{A}\right)^{-1}$
is \cite{book:538623}
\[
\left(\mathbf{I}+\mathbf{A}\right)^{-1}=\mathbf{I}+\sum_{n=1}^{\infty}\left(-1\right)^{n}\mathbf{A}^{n}.
\]
The inversion is possible when the absolute value of all the eigenvalues
of $\mathbf{A}$ is less than one. Considering now $\mathbf{A}=\mathbf{e}^{-\mathbf{\mathbf{\Gamma}}x}\bm{\rho}_{\mathbf{L}}^{\mathbf{M}}\mathbf{e}^{-\mathbf{\mathbf{\Gamma}}x}$,
its eigenvalues can in principle be greater than one since any component
of $\bm{\rho}_{\mathbf{L}}^{\mathbf{M}}$ can be greater than one
\cite{6312388}. However, we suppose that the magnitude of the eigenvalues
of $\bm{\rho}_{\mathbf{L}}^{\mathbf{M}}$ is sufficiently damped by
the exponential matrices, which is normally the case in PLNs. The
input admittance matrix \eqref{eq:yinsingle} can be therefore rewritten
as  
\begin{equation}
\mathbf{Y}_{\mathbf{in}}=\mathbf{T}\left[\mathbf{I}+2\sum_{n=1}^{\infty}\left(\mathbf{e}^{-\mathbf{\Gamma}\ell_{1}}\bm{\rho}_{\mathbf{L}}^{\mathbf{M}}\mathbf{e}^{-\mathbf{\Gamma}\ell_{1}}\right)^{n}\right]\mathbf{T}^{-1}\mathbf{Y}_{\mathbf{C}}\label{eq:yinsingleexpanded}
\end{equation}
and \eqref{eq:pinsingle} can be rewritten as 
\begin{multline}
\bm{\rho}_{\mathbf{in}}=\mathbf{N}\mathbf{T}\bm{\rho}_{\mathbf{G}}^{\mathbf{M}}\mathbf{T}^{-1}\mathbf{N}^{-1}+\mathbf{N}\mathbf{T}\Bigg[\left(\mathbf{I}-\bm{\rho}_{\mathbf{G}}^{\mathbf{M}}\bm{\rho}_{\mathbf{G}}^{\mathbf{M}}\right)\mathbf{e}^{-\mathbf{\Gamma}\ell_{1}}\\
\bm{\rho}_{\mathbf{L}}^{\mathbf{M}}\mathbf{e}^{-\mathbf{\Gamma}\ell_{1}}\sum_{n=\text{0}}^{\infty}\left(-1\right)^{n}\left(\bm{\rho}_{\mathbf{G}}^{\mathbf{M}}\mathbf{e}^{-\mathbf{\Gamma}\ell_{1}}\bm{\rho}_{\mathbf{L}}^{\mathbf{M}}\mathbf{e}^{-\mathbf{\Gamma}\ell_{1}}\right)^{n}\Bigg]\mathbf{T}^{-1}\mathbf{N}^{-1}.\label{eq:pinsingleexpanded}
\end{multline}
\eqref{eq:yinsingleexpanded} and \eqref{eq:pinsingleexpanded} are
composed by the sum of three elements: a constant w.r.t. $\ell_{1}$,
a first exponential function depending on $2\ell_{1}$, and an infinite
series of exponentials depending on multiples of this distance. We
now make use of the matrix inverse Fourier transform defined for any
\textbf{X} matrix as 
\begin{equation}
\mathbf{x}(t)=\int_{-\infty}^{+\infty}\mathbf{X}(f)e^{j2\pi ft}df,
\end{equation}
and apply it to \eqref{eq:yinsingleexpanded} and \eqref{eq:pinsingleexpanded}.
The resulting multidimensional time traces are 
\begin{multline}
\mathbf{y}_{\mathbf{in}}(t)=\Bigg[\bm{\delta}\left(t\right)\ast\mathbf{e_{0}}\ast\bm{\delta}\left(t\right)\\
+2\sum_{n=1}^{\infty}\bm{\delta}\left(t-2n\mathbf{t}_{1}\right)\ast\mathbf{e_{n}}\ast\bm{\delta}\left(t-2n\mathbf{t}_{1}\right)\Bigg]\ast\mathbf{y}_{\mathbf{C}_{1}}(t)\label{eq:yonetime}
\end{multline}
and 
\begin{multline}
\bm{\rho}_{\mathbf{in}}(t)=\bm{\delta}\left(t\right)\ast\mathbf{f_{0}}\ast\bm{\delta}\left(t\right)+\sum_{n=1}^{\infty}\bm{\delta}\left(t-2\mathbf{t}_{1}\right)\ast\mathbf{f_{n}}\ast\bm{\delta}\left(t-2n\mathbf{t}_{1}\right)\label{eq:ponetime}
\end{multline}
where $\ast$ denotes the matrix-wise convolution operation, $\mathbf{e_{n}}$
and $\mathbf{f_{n}}$ denote functions that depend on time but not
on the position of the nodes, $\bm{\delta}\left(t\right)$ is a diagonal
matrix whose elements are delta functions and $\mathbf{t}_{1}$ is
a vector of propagation times (the propagation velocities over the
different conductor pairs might be different) between the signal source
and the load. We point out that $\mathbf{e_{0}}=\mathbf{I}$ and $\mathbf{f_{0}}=\mathbf{N}\bm{\rho}_{\mathbf{G}}^{\mathbf{}}\mathbf{N}^{-1}$,
so the envelope of the first peak depends only on $\mathbf{y}_{\mathbf{C}_{1}}$
and its eventual mismatch with $\mathbf{Y}_{\mathbf{R}}$. 

If we suppose constant propagation velocity $\mathbf{v}$ and $\mathbf{\Gamma}(f)=\mathbf{\Gamma}_{1}f$,
where $f$ is the frequency, then $\mathbf{t}_{1}=\mathbf{\Gamma}_{1}\mathbf{\ell}_{1}$.
This case applies for ideal TLs, but in practical scenarios $\mathbf{\Gamma}$
might not have a direct proportionality to $f$ \cite{tonello2012a}.
More in general, we can write 
\begin{equation}
\mathbf{t}_{1}=\mathbf{g}\left(\mathbf{\Gamma}_{1},\mathbf{\ell}_{1}\right),\label{eq:t1}
\end{equation}
where $\mathbf{g}$ is a generic function. 

We consider now a complex MTL network with a tree topology (no loops)
made by N nodes and N-1 branches. The $m$th branch is characterized
by its length $\ell_{m}$ and propagation matrix $\mathbf{\mathbf{\Gamma_{m}}}$.
The input admittance matrix at the generator side can be expressed
as (derivation in Appendix \ref{sec:Derivation-of-the})
\begin{multline}
\mathbf{Y}_{\mathbf{in}}=\mathbf{T_{1}}\Bigg[\mathbf{I}+2\sum_{n=1}^{N}\prod_{m=1}^{n}\mathbf{A_{m}}\mathbf{e}^{-\mathbf{\mathbf{\Gamma_{m}}}\ell_{m}}\\
\prod_{m=0}^{n-1}\mathbf{B_{n-m}}\mathbf{e}^{-\mathbf{\mathbf{\Gamma_{n-m}}}\ell_{n-m}}+2\sum_{n=N+1}^{\infty}\dots\Bigg]\mathbf{T_{1}}^{-1}\mathbf{Y}_{\mathbf{C}_{1}}\label{eq:Yingeneral}
\end{multline}
and the input reflection coefficient matrix becomes
\begin{multline}
\bm{\rho}_{\mathbf{in}}=\mathbf{N_{1}}\mathbf{T_{1}}\Bigg[\bm{\rho}_{\mathbf{G_{1}}}^{\mathbf{M}}+\sum_{n=1}^{N}\prod_{m=1}^{n}\mathbf{C_{m}}\mathbf{e}^{-\mathbf{\mathbf{\Gamma_{m}}}\ell_{m}}\\
\prod_{m=0}^{n-1}\mathbf{D_{n-m}}\mathbf{e}^{-\mathbf{\mathbf{\Gamma_{n-m}}}\ell_{n-m}}+\sum_{n=N+1}^{\infty}\dots\Bigg]\mathbf{T_{1}}^{-1}\mathbf{N_{1}}^{-1}\label{eq:pingeneral}
\end{multline}
where the subscript $_{1}$ refers to the line segment to which the
signal source is branched and $\mathbf{A_{m}}$, $\mathbf{B_{m}}$,
$\mathbf{C_{m}}$, $\mathbf{D_{m}}$ are frequency dependent functions.
We represented \eqref{eq:Yingeneral} and \eqref{eq:pingeneral} in
this form to highlight the fact that $N$ components of the sum are
a direct function of double the distance of each of the $N$ nodes
of the network from the signal generator. The sum proceeds to infinity
with components that are functions of the first $N$ terms. Finally,
both $\mathbf{Y_{in}}$ and $\bm{\rho}_{\mathbf{in}}$ can be written
in the more compact form 
\begin{equation}
\mathbf{Y}_{\mathbf{in}}=\mathbf{T_{1}}\Bigg[\sum_{n=1}^{\infty}\mathbf{E_{n}}\mathbf{e}^{-\mathbf{\mathbf{\Gamma eq_{n}}}\ell eq_{n}}\Bigg]\mathbf{T_{1}}^{-1}\mathbf{Y}_{\mathbf{C}_{1}}\label{eq:Ymoregeneral}
\end{equation}
and
\begin{equation}
\bm{\rho}_{\mathbf{in}}=\mathbf{N_{1}}\mathbf{T_{1}}\Bigg[\bm{\rho}_{\mathbf{G_{1}}}^{\mathbf{M}}+\sum_{n=1}^{\infty}\mathbf{F_{n}}\mathbf{e}^{-\mathbf{\mathbf{\Gamma eq_{n}}}\ell eq_{n}}\Bigg]\mathbf{T_{1}}^{-1}\mathbf{N_{1}}^{-1},\label{eq:rhomoregeneral}
\end{equation}
where $\mathbf{E_{n}}$, $\mathbf{F_{n}}$ are frequency dependent
functions and the index $n$ does not anymore refer to a branch but
to a path instead. Consequently, $\ell eq$ and $\mathbf{\Gamma eq}$
are the length and the propagation constant of a specific propagation
path ($\mathbf{\Gamma eq}$ is not a diagonal matrix as \textbf{$\mathbf{\Gamma}$}).
In this form, \eqref{eq:Ymoregeneral} and \eqref{eq:rhomoregeneral}
can be manipulated by standard frequency analysis tools to retrieve
information about the parameters $\ell eq$, $\mathbf{\Gamma eq}$,
$\mathbf{E_{n}}$ and $\mathbf{F_{n}}$\cite{stoica2005spectral}.

We now transform the aforementioned equations to the time domain by
performing a multidimensional Fourier transform, which results in
\begin{multline}
\mathbf{y}_{\mathbf{in}}(t)=\bm{\delta}\left(t\right)\ast\mathbf{y_{C_{1}}}\ast\bm{\delta}\left(t\right)\\
+2\sum_{n=1}^{N}\bm{\delta}\left(t-2\mathbf{t}_{n}\right)\ast\mathbf{e_{n}}\ast\bm{\delta}\left(t-2\mathbf{t}_{n}\right)+2\sum_{n=N+1}^{\infty}\dots\label{eq:ytime}
\end{multline}
\begin{multline}
\bm{r}_{\mathbf{in}}(t)=\bm{\delta}\left(t\right)\ast\bm{r}_{\mathbf{G_{1}}}\ast\bm{\delta}\left(t\right)\\
+\sum_{n=1}^{N}\bm{\delta}\left(t-2\mathbf{t}_{n}\right)\ast\mathbf{f_{n}}\ast\bm{\delta}\left(t-2\mathbf{t}_{n}\right)+\sum_{n=N+1}^{\infty}\dots\label{eq:ptime}
\end{multline}
where $t_{n}$ is the flight time of the signal from the source to
the $n$-th node, and $\mathbf{y_{C_{1}}}$ , $\mathbf{r_{in}}$ and
$\bm{r}_{\mathbf{G_{1}}}$ are the inverse Fourier transforms of $\mathbf{Y_{C_{1}}}$,
$\bm{\rho}_{\mathbf{in}}$ and $\mathbf{N_{1}}\mathbf{T_{1}}\bm{\rho}_{\mathbf{G_{1}}}^{\mathbf{M}}\mathbf{T_{1}}^{-1}\mathbf{N_{1}}^{-1}$
respectively. \eqref{eq:ytime} and \eqref{eq:ptime} show that the
time domain reflectometric response of a MTL comprises a first signal
starting at $t=0$ which is generated by the impedance mismatch of
the first line segment. This is followed by a series of smoothed peaks
at $2\mathbf{t}_{n}$, $n\in1\dots N$ that identify the presence
of the $N$ nodes of the network. Another infinite series of peaks
is summed up in the time trace, which are located at time instants
that are linear combinations of the first $N$.

This information is commonly used when the topology of the network
is not known. In fact, by converting time to distance, a reflectometric
measurement gives information about the distance of all the network
nodes from the measurement point. Different algorithms can be contextually
exploited to merge this information and estimate the topology of the
network (see for example \cite{neusphd,1007375}).

Moreover, since the reflectometric frequency responses \eqref{eq:Yingeneral}
and \eqref{eq:pingeneral} are written as sums of damped exponentials,
parametric models can be applied in special conditions in order to
enhance the accuracy in the estimation of the position of the peaks
(see for example \cite{ahmed2012topology2}).

\subsection{Analysis of the end-to-end propagation\label{sub:End-to-end}}

The channel transfer function (CTF) $\mathbf{H}$ of a single MTL
cable with length $\ell_{1}$ and a load $\mathbf{Y_{L}}$ at the
end can be written as \cite{versolatto2011an}
\begin{multline}
\mathbf{H}=\mathbf{Y_{C}^{-1}T}\left(\mathbf{I}-\bm{\rho}_{\mathbf{L}}\right)\left(\mathbf{I}-\mathbf{e}^{-2\mathbf{\Gamma}\ell_{1}}\bm{\rho}_{\mathbf{L}}\right)^{-1}\mathbf{e}^{-\mathbf{\Gamma}\ell_{1}}\mathbf{T}^{-1}\mathbf{Y_{C}}\label{eq:Hsinple}
\end{multline}
where the equivalence holds because $\mathbf{\Gamma}$ is a diagonal
matrix. The voltage on the load results in 
\begin{equation}
\mathbf{V_{L}}=\mathbf{HV_{source}}.
\end{equation}
Using the same approach of the previous section, \eqref{eq:Hsinple}
can be rewritten as
\begin{multline}
\mathbf{H}=\mathbf{Y_{C}^{-1}T}\left(\mathbf{I}-\bm{\rho}_{\mathbf{L}}\right)\Bigg[\sum_{n=0}^{\infty}\left(\mathbf{e}^{-2\mathbf{\Gamma}\ell_{1}}\bm{\rho}_{\mathbf{L}}\right)^{n}\Bigg]\mathbf{e}^{-\mathbf{\Gamma}\ell_{1}}\mathbf{T}^{-1}\mathbf{Y_{C}}.\label{eq:H1}
\end{multline}
The corresponding time domain multidimensional trace is 
\begin{multline}
\mathbf{h}(t)=\mathbf{y}_{\mathbf{C}_{1}}^{-1}\ast\sum_{n=0}^{\infty}\Bigg[\bm{\delta}\left(t-2(n+1)\mathbf{t}_{1}\right)\ast\mathbf{a_{n}}\\
*\bm{\delta}\left(t-2(n+1)\mathbf{t}_{1}\right)\Bigg]\ast\mathbf{y}_{\mathbf{C}_{1}}
\end{multline}
where $\mathbf{a_{n}}$ is a time dependant matrix.

When a complex MTL network made by several different branches is considered,
the chain rule of the CTF can be applied, so that the voltage on the
load of the receiving end reads
\begin{equation}
\mathbf{V_{L}}=\mathbf{H_{tot}}\mathbf{V_{source}}.
\end{equation}
where 
\begin{multline}
\mathbf{H_{tot}}=\prod_{n=1}^{N}\mathbf{H_{n}}=\prod_{n=1}^{N}\mathbf{Y_{C_{n}}^{-1}T_{n}}\left(\mathbf{I}-\bm{\rho}_{\mathbf{L_{n}}}\right)\\
\Bigg[\sum_{m=0}^{\infty}\left(\mathbf{e}^{-2\mathbf{\Gamma_{n}}\ell_{n}}\bm{\rho}_{\mathbf{L}_{n}}\right)^{m}\Bigg]\mathbf{e}^{-\mathbf{\Gamma}\ell_{n}}\mathbf{T_{n}}^{-1}\mathbf{Y_{C_{n}}}\label{eq:Htotal}
\end{multline}
and $\bm{\rho}_{\mathbf{L_{n}}}$ is the equivalent reflection coefficient
matrix at the end of each line segment. We remark that every $\mathbf{H_{n}}$
is derived using the carry-back procedure presented in \cite{tonello2011bottomup}.
Finally, similarly to \eqref{eq:Ymoregeneral} and \eqref{eq:rhomoregeneral},
$\mathbf{H_{tot}}$ can be more in general written as
\begin{equation}
\mathbf{H_{tot}}=\sum_{n=1}^{\infty}\mathbf{A_{n}}\mathbf{e}^{-\mathbf{\mathbf{\Gamma eq_{n}}}\ell_{n}},\label{eq:Hmoregeneral}
\end{equation}
which allows thorough analysis of the transfer function using spectral
analysis algorithms.

The product of exponential sequences in \eqref{eq:Htotal} becomes
a convolution of delta functions in time domain. The time domain
transfer function acquires the following form:
\begin{multline}
\mathbf{h_{tot}}(t)=\sum_{m=1}^{\infty}\bm{\delta}\left(t-\sum_{n=1}^{N}p_{n,m}\mathbf{t}_{n}\right)\ast\mathbf{a_{n,m}}\\
\ast\bm{\delta}\left(t-\sum_{n=1}^{N}p_{n,m}\mathbf{t}_{n}\right),\label{eq:htotal}
\end{multline}
where $p_{n}$ is the number of times the signal has traveled through
the $n$-th line segment. \eqref{eq:htotal} represents an infinite
series of smoothed pulses where the delay of each pulse is due to
the total number of line segments the signal travels across before
reaching the receiver. The number of possible paths is of course infinite,
but it is important to remark that their spacing is, as for the reflectometry
case, a function of the length of all the line segments in the network.
Therefore, we might suppose that also end-to-end signals can be used
with the aim of topology estimation. That is however not possible,
due to the following
\begin{thm}[Time Domain Wide-Sense Symmetry]
\label{thm:TDS}When considering end-to-end propagation in a passive
medium from point A to point B, the distance between each smoothed
peak in the channel impulse response is the same as in the case of
propagation from point B to point A, irrespective of the complexity
of the scattering elements, their reflection coefficients, the impedance
at the transmitter and at the receiver.
\end{thm}
An intuitive demonstration of Theorem \ref{thm:TDS} is given considering
that the system is a two port passive network and, as all these kind
of networks, it is reciprocal. A thorougher demonstration is given
in Appendix \ref{sec:Proof-of-Theorem}. We point out that the Theorem
does not specify anything about the amplitude and the shape of the
peaks. In fact, they are in general different when considering the
A-to-B or B-to-A propagation paths. Because of Theorem \ref{thm:TDS},
it is not possible to know whether a peak in the measured $\mathbf{h_{tot}}(t)$
identifies a node close to the near or the far end of the communication
link. Hence, end-to-end communication cannot be used to reconstruct
the topology of a network univocally. When applied to the case of
anomaly estimation, the same principle applies. End-to-end communication
can identify the distance of the anomaly, but there is always an ambiguity
whether it is the distance from the transmitter or the receiver.

\section{Effect of anomalies on the signal propagation\label{sec:Effect-of-anomalies}}

In this section, we present an analysis of the effect of electrical
anomalies caused by faults or aged cables on the propagation of the
signals in a PLN. 

\begin{figure}[tb]
\begin{centering}
\subfloat[Lumped fault]{\begin{centering}
\includegraphics[width=0.48\columnwidth]{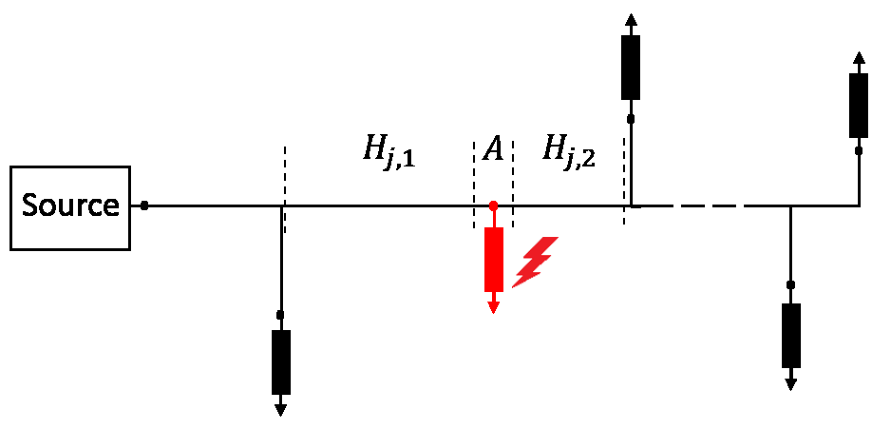}
\par\end{centering}

}\subfloat[Termination impedance change]{\begin{centering}
\includegraphics[width=0.48\columnwidth]{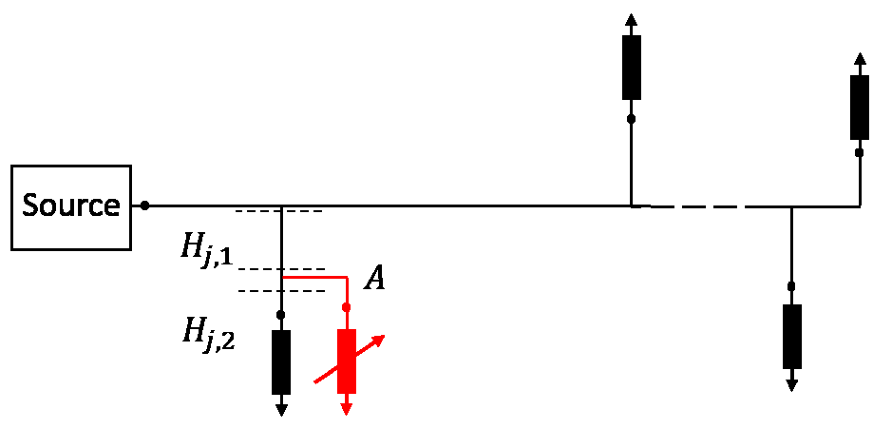}
\par\end{centering}

}
\par\end{centering}

\begin{centering}
\subfloat[Distributed fault]{\begin{centering}
\includegraphics[width=0.48\columnwidth]{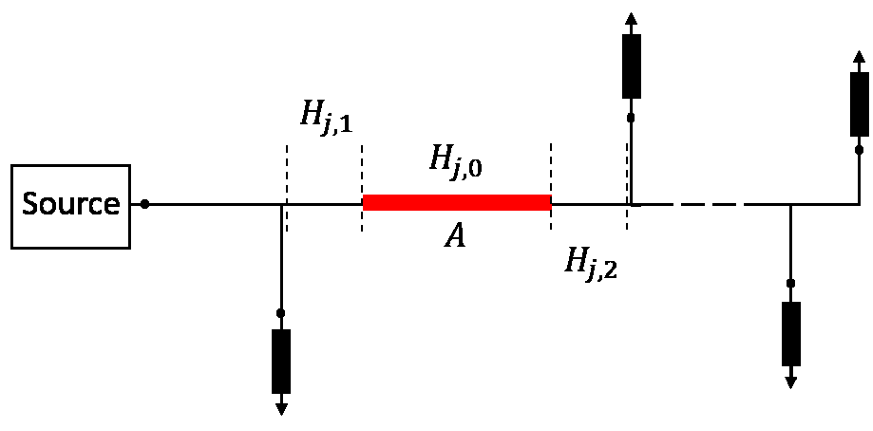}
\par\end{centering}

}
\par\end{centering}

\caption{Sketch of the electrical equivalent of different anomalies in a distribution
network.\label{fig:Sketch-fault}}
\end{figure}

In a broad sense, an anomaly is a modification in the expected behavior
of a system. In the case of SGs we identify three main categories
of anomalies: concentrated faults, distributed faults and termination
impedance changes. Both localized faults and impedance changes can
be schematically represented as a Thevenin or Norton equivalent circuit
that is superimposed at the fault location to the previously known
system (see Fig. \ref{fig:Sketch-fault}). In the literature, these
circuits are normally reduced to a lumped impedance \cite{sgc2017}
or, equivalently, to a voltage or current generator \cite{7878929}.
On the other side, distributed faults like damaged cables, can only
be represented by a cable section with modified parameters. 
\begin{figure}[tb]
\centering{}\includegraphics[width=0.9\columnwidth]{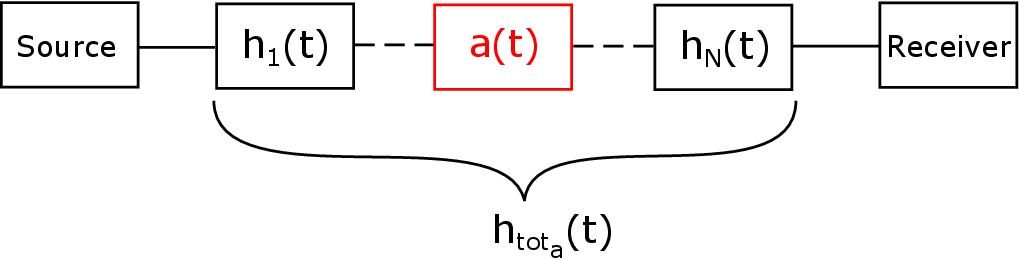}\caption{Model of the anomaly using the chain representation.\label{fig:anomaly-chain}}
\end{figure}

\subsection{General models of the anomalies\label{sub:General-models-of}}

The effect of such anomalies can be represented by an extra transfer
function block $\mathbf{A}$ inserted in the network chain model at
the position of the anomaly occurrence (see Figure \ref{fig:anomaly-chain}).
\begin{figure}[tb]
\begin{centering}
\includegraphics[width=1\columnwidth]{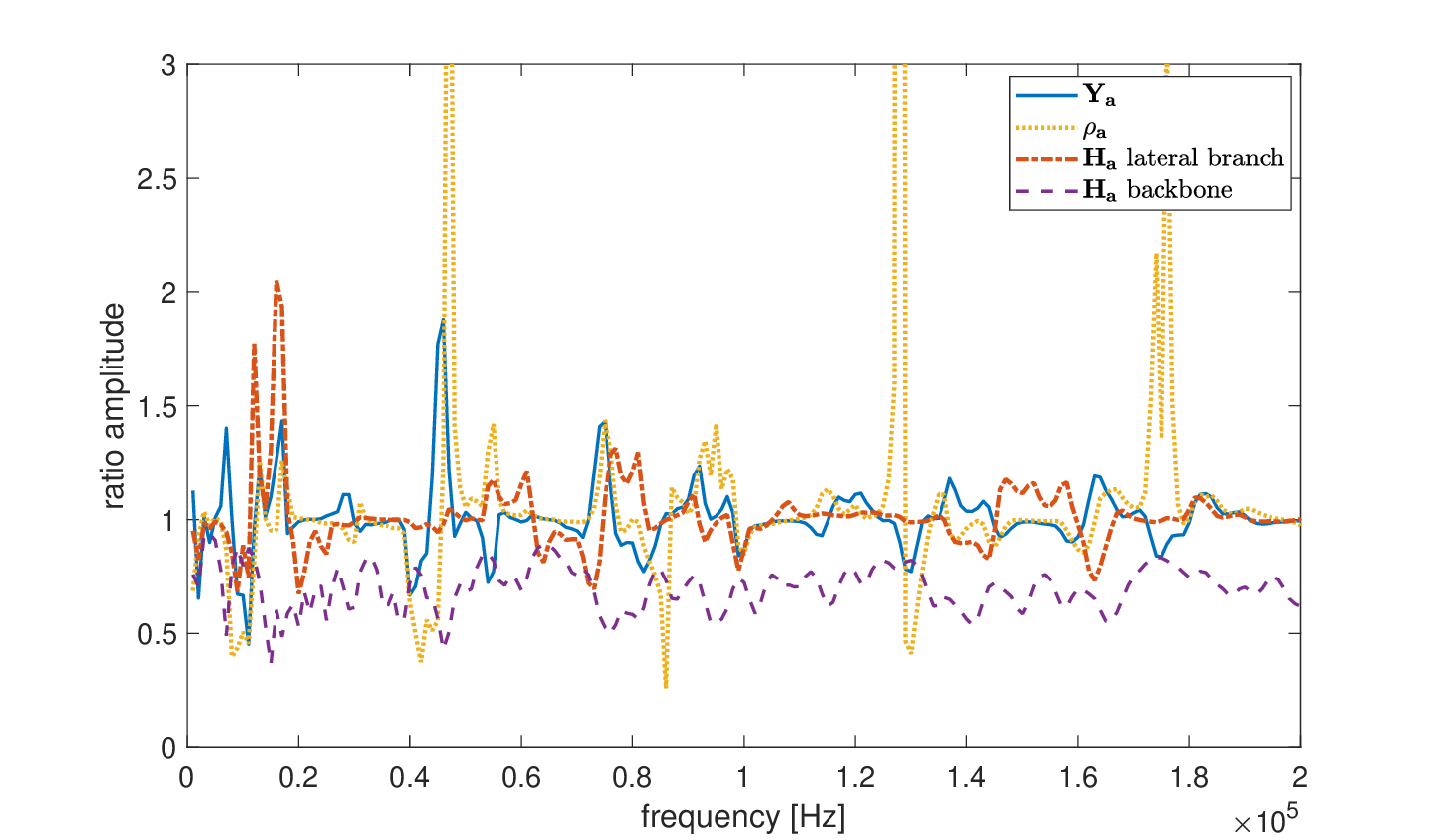}
\par\end{centering}

\caption{Example of typical values obtained using the chain model. Two examples
of $\mathbf{H_{a}}$ are given: when the fault occurs on the backbone
and when it occurs on a lateral branch.\label{fig:Example-of-typical} }

\end{figure}

If we write the transfer function of branch $j$ as $\mathbf{H_{j}}=\mathbf{H_{j,1}}\mathbf{H_{j,0}}\mathbf{H_{j,2}}$,
in the presence of an anomaly on branch $j$, \eqref{eq:Htotal} becomes
\begin{equation}
\mathbf{H_{tot_{a}}}=\left(\prod_{n=1}^{j-1}\mathbf{\tilde{H}_{n}}\right)\mathbf{\tilde{H}_{j,1}}\mathbf{A}\mathbf{H_{j,2}}\left(\prod_{n=j+1}^{N}\mathbf{H_{n}}\right).\label{eq:Htota}
\end{equation}
Here $\mathbf{A}$ replaces $\mathbf{H_{j,0}}$, and $\mathbf{\tilde{H}_{j,1}}$
is equal to $\mathbf{H_{j,1}}$  where the load admittance is $\mathbf{Y_{in_{A}}}$
instead of $\mathbf{Y_{in_{j,2}}}$. The modification of the load
admittance is consequently propagated through all the $\mathbf{\tilde{H}_{n}}$,
$n=1\dots j-1$. When a load impedance change or a concentrated fault
is considered, $\mathbf{H_{j,0}}$ is an identity matrix and $\mathbf{A}$
is the transfer function of an impedance matrix. When a distributed
fault is considered, $\mathbf{H_{j,0}}$ is the transfer function
of the healthy cable and $\mathbf{A}$ is the transfer function of
the degraded cable. If we now want to obtain the CTF variation $\Delta_{ch}^{H}$,
which represents the compound effect of the anomaly on the network,
we can compute the product 
\begin{equation}
\mathbf{H_{tot_{a}}}\mathbf{H_{tot}}^{-1}=\left(\prod_{n=1}^{j,1}\mathbf{\tilde{H}_{n}}\right)\mathbf{A}\left(\prod_{n=0}^{(j,1)-1}\mathbf{H_{(j,1)-n}}^{-1}\right)=\Delta_{ch}^{H}.\label{eq:Ha}
\end{equation}
Since the product from $j+1$ to $N$ is the same in \eqref{eq:Htotal}
and \eqref{eq:Htota}, the $N-j$ branches that are closest to the
receiver are canceled out in \eqref{eq:Ha}. Hence, the system is
reduced to a network whose closest node to the receiver is the one
generated by the anomaly. The first relevant exponential in $\Delta_{ch}^{H}$
is $\mathbf{e}^{-2\mathbf{\Gamma eq}\ell eq_{a}}$, where $\ell eq_{a}$
is the distance of the beginning of the anomaly from the closest node
to the anomaly. We point out that, if the anomaly is a fault and it
occurs along the backbone, i.e. the direct path, between the transmitter
and the receiver, it creates a further propagation path for the direct
propagation wave, which in turn results in the reduction of the CTF.
Conversely, when the anomaly occurs on a lateral branch, it does only
affect secondary propagating waves, resulting on average in no signal
loss (see Fig. \eqref{fig:Example-of-typical}).

Considering now the case of the input admittance, starting from \eqref{eq:Ymoregeneral}
we compute the admittance variation $\Delta_{ch}^{Y}$ as
\begin{multline}
\Delta_{ch}^{Y}=\mathbf{Y}_{\mathbf{in_{a}}}\mathbf{Y_{in}}^{-1}=\\
\mathbf{T_{1}}\Bigg[\sum_{m=1}^{\infty}\mathbf{E_{m}}\mathbf{e}^{-\mathbf{\mathbf{\Gamma eq_{m}}}\ell eq_{m}}+\sum_{n=1}^{\infty}\mathbf{F_{n}}\mathbf{e}^{-\mathbf{\mathbf{\Gamma eq_{n}}}\ell eq_{n}}\Bigg]\\
\Bigg[\sum_{p=1}^{\infty}\mathbf{E_{p}}\mathbf{e}^{-\mathbf{\mathbf{\Gamma eq_{p}}}\ell eq_{p}}\Bigg]^{-1}\mathbf{T_{1}}^{-1},\label{eq:Ya}
\end{multline}
where the sum over $n$ refers to the new paths caused by the anomaly.
We see that, since the signal is transmitted from and returns to the
same point, all the exponentials due to the nodes between the sensing
point and the anomaly location erase each other, and the first significant
exponential is the one for $n=1$. $\ell eq_{1}$ is actually the
distance from the anomaly from the sensing point. 

Moving to time domain, \eqref{eq:Ha} becomes
\begin{multline}
\partial_{ch}^{H}(t)=\bm{\delta}\left(t\right)\ast\mathbf{a_{0}}\ast\bm{\delta}\left(t\right)+\sum_{m=1}^{\infty}\bm{\delta}\left(t-\sum_{n=1}^{j}p_{n}\mathbf{t}_{n}\right)\\
\ast\mathbf{a_{n}}\ast\bm{\delta}\left(t-\sum_{n=1}^{j}p_{n}\mathbf{t}_{n}\right),\label{eq:ha}
\end{multline}
where the peak at 0 is due to the bias introduced by the multiplication
in \eqref{eq:Ha}, and the second peak is at $\mathbf{t_{a}}$, associated
to the distance $\ell_{a}$ of the anomaly from the receiver. A similar
equation can be written starting from \eqref{eq:Ya}, but in this
case $\mathbf{t_{a}}$ univocally identifies the distance of the anomaly
from the sensing point. This information is in general more encripted
in the peak series of $\partial_{ch}(t)$. If the topology of the
network is a-priori known, the position of the peaks can be related
to a specific position in the network. More on this is discussed in
\cite{SGSII}.

$\bm{\rho}_{\mathbf{in}}$ deserves a separate treatment. In fact,
its values can be greater or lower than $0$ and are often close to
it. This means that at a certain frequency $\bm{\rho}_{\mathbf{in}}$
can be greater than $0$ and $\bm{\rho}_{\mathbf{in_{a}}}$can be
lower. This might yield very high absolute values of $\bm{\rho}_{\mathbf{a}}$
(see Fig. \ref{fig:Example-of-typical}), but results in a distortion
of the information about the anomaly. The resulting $\mathbf{r_{a}}(t)$
is in general rather jagged and difficult to analyze. For this reason
it is not advisable to use the chain model when sensing $\bm{\rho}_{\mathbf{in}}$.

\begin{figure}[tb]
\centering{}\includegraphics[width=0.7\columnwidth]{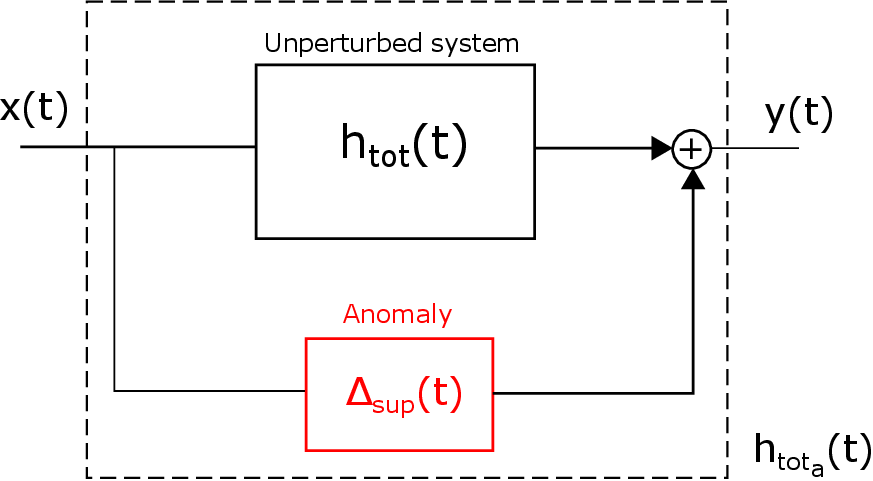}\caption{Model of the anomaly using the superposition of the effects.\label{fig:Model-of-the}}
\end{figure}

If we now consider again \eqref{eq:Ymoregeneral}, \eqref{eq:rhomoregeneral}
and \eqref{eq:Hmoregeneral}, the anomaly can also be modeled using
the superposition of the effects as an independent transfer function
$\mathbf{A_{S}}$ that adds to the unperturbed system (see Fig. \eqref{fig:Model-of-the}).
This is the same model used in the radar literature for the detection
of moving objects in cluttered environments \cite{4200703,soltanalian,abboud}.
Relying on this model, the effect of the anomaly on the system can
be derived by subtracting $\mathbf{H_{tot}}$ from $\mathbf{H_{tot_{a}}}$
. Since $\mathbf{H_{tot_{a}}}=\Delta_{ch}^{H}\mathbf{H_{tot}}$, and
$\Delta_{ch}^{H}$ can be written as 
 
\begin{equation}
\Delta_{ch}^{H}=\sum_{m=1}^{\infty}\mathbf{B_{m}}\mathbf{e}^{-\mathbf{\mathbf{\Gamma eq_{m}}}\ell eq_{m}},\label{eq:Hamoregeneral}
\end{equation}
we have

\begin{equation}
\Delta_{sup}^{H}=\left(\Delta_{ch}^{H}-\mathbf{I}\right)\mathbf{H_{tot}}=\sum_{k=1}^{\infty}\mathbf{C_{k}}\mathbf{e}^{-\mathbf{\mathbf{\Gamma eq_{k}}}\ell eq_{k}}.\label{eq:As}
\end{equation}
The reflectometry case is similarly derived. We see from \eqref{eq:As}
that all the exponential functions of \eqref{eq:Hmoregeneral} are
still present, but their amplitude is modified by the effect of \textbf{$\Delta_{ch}^{H}$}.
All the exponential functions relative to the position of the anomaly
are also added up in the sum. \eqref{eq:As} can be further normalized
by multiplication with $\mathbf{H_{tot}}^{-1}$, so that 
\begin{equation}
\Delta_{sup_{N}}^{H}=\Delta_{sup}^{H}\mathbf{H_{tot}}^{-1}=\Delta_{ch}^{H}-1,\label{eq:Asn}
\end{equation}
which is basically the percentage variation of the measured CTF due
to the anomaly. In time domain, \eqref{eq:As} becomes
\begin{equation}
\partial_{sup}^{H}(t)=\left(\partial_{ch}^{H}(t)-\bm{\delta}\left(t\right)\right)\ast\mathbf{h_{tot}}(t),\label{eq:as}
\end{equation}
which includes all the peaks of $\mathbf{h_{tot}}(t)$ plus the new
peaks generated by the fault. This can be also intuitively derived
by considering that in the anomalous condition the test signal, which
travels from the transmitter to a far receiver, is already modified
by the anomaly before reaching the receiver. Therefore, already the
first peak at the receiver is modified by the anomaly with respect
to the unperturbed situation. 

The reflectometry case is rather different when considering the superposition
model. In fact, the test signal is sent and returns to the same point.
The effect of an anomaly would therefore not influence the transitory
until the transmitter signal has reached the anomaly. This means that
all the possible peaks that occur before the anomaly are canceled
out when computing $\mathbf{Y}_{\mathbf{in}}-\mathbf{Y}_{\mathbf{in_{a}}}$. 

Hence, although the chain model and the superposition model are in
principle different, they provide the same information about the effect
of an anomaly, when considering the reflectometry case. In the end-to-end
transmission case instead, the two models yield different results.
In the chain model, all the first part of the transmission is canceled
out, so that the first peak in time domain identifies the distance
of the anomaly from the closest node, while the superposition model
features a trace with the same peak positions of the unperturbed situation,
plus a series of peaks due to the anomaly. 

We see here that, although both the end-to-end and reflectometric
sensing approaches can be used to detect the presence of an anomaly,
the location of the same is easier using the reflectometric approach,
since the first peak of $\partial_{ch}^{Y}(t)$ or $\partial_{sup}^{Y}(t)$
already provides the distance of the anomaly from the sensing point.

\subsection{Effect of concentrated faults and load impedance changes}

A fault can be described electrically as a concentrated impedance
branched to the network at the point where it occurs. Many models
have been described in the literature (see \cite{amhifatl,6311450}
and references therein) to best describe with lumped elements the
dynamic evolution of different type of faults. In this paper, we will
more generically consider faults with generic spectra. 

A load impedance change can be due to two reasons: a stable variation
of the load or a cyclic impedance variation due to the presence of
active power converters. 

\begin{figure}[tb]
\begin{centering}
\subfloat[Frequency domain]{\begin{centering}
\includegraphics[width=1\columnwidth]{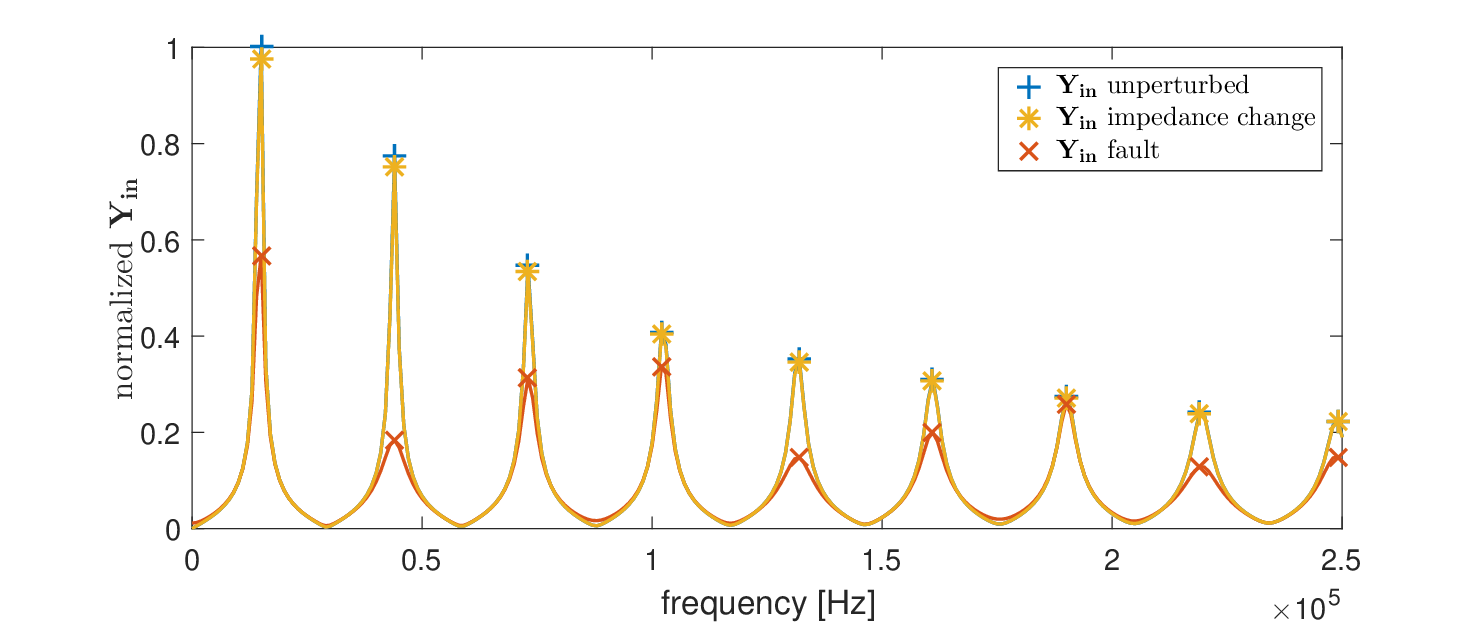}
\par\end{centering}

}
\par\end{centering}

\centering{}\subfloat[Time domain]{\begin{centering}
\includegraphics[width=1\columnwidth]{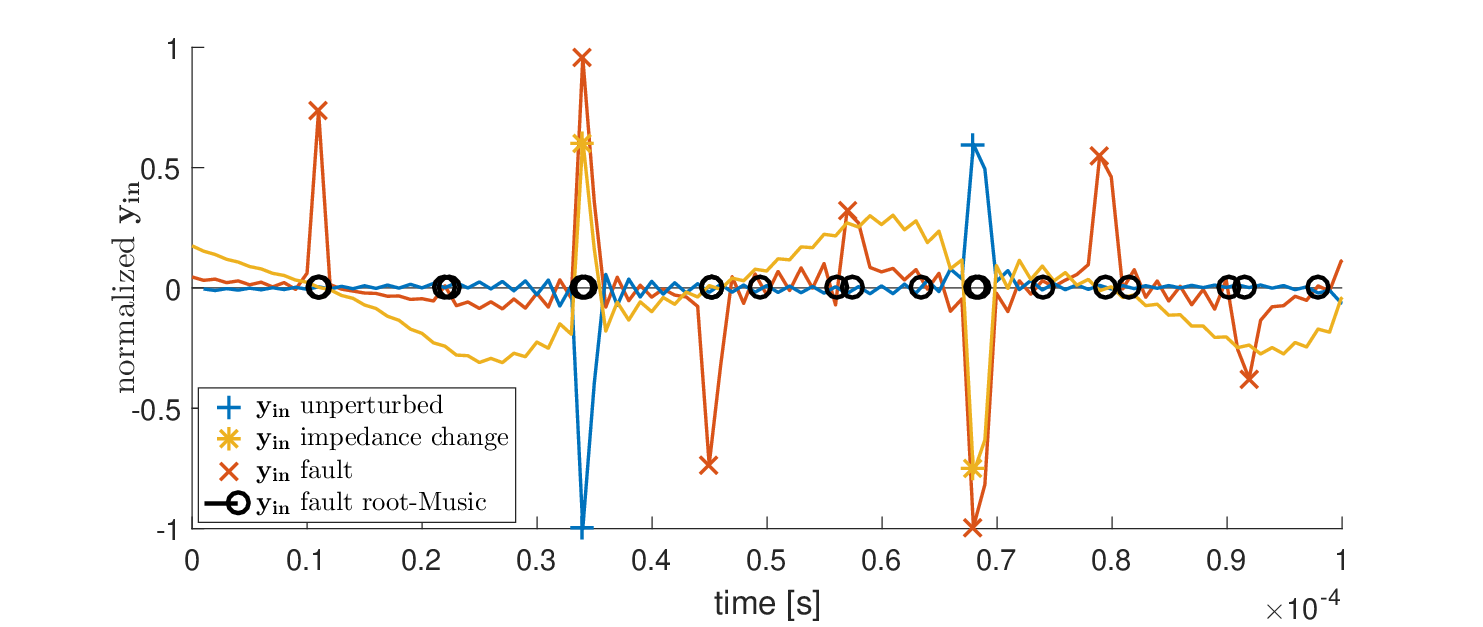}
\par\end{centering}

}\caption{$\mathbf{Y_{in}}$ simulated in presence of a fault and a load impedance
change in frequency (a) and time (b) domain.\label{fig:-simulated-in}}
\end{figure}

From an electrical point of view, faults and impedance changes produce
the same effect on the channel, since they are both concentrated and
introduce the additional transfer function $\mathbf{A}$ of \eqref{eq:Htota}
in the system. However, conversely from an impedance change, a fault
modifies the topological structure of the network by introducing a
new node with the fault impedance as a load (see Fig. \ref{fig:Sketch-fault}).
The consequence is that an impedance change is always localized at
the same position of a previously known peak, and the same holds for
all the following peaks of $\mathbf{H_{a}}$. On the other hand, all
the peaks generated by a fault are new with respect to the unperturbed
system. In Figure \ref{fig:-simulated-in}, the difference between
a fault and a load impedance change is shown in the simple case of
a single transmission line. Impedance measurements and the superposition
model are used to derive the plots. The frequency domain plot shows
that a load impedance change simply modifies the amplitude of $\mathbf{Y_{in}}$,
especially at low frequencies. Conversely, a fault introduces a new
oscillating mode that sums up to the exponential decaying trend, as
explained in Section \ref{sub:General-models-of}. The time domain
plot confirms that faults actually introduce new peaks with respect
to the unperturbed situation, while load impedance changes simply
modify their amplitude. 

\subsection{Effect of distributed faults}

Distributed faults involve the damage of an extended section of a
cable. Such faults can be due to exposure to bad weather conditions,
water leakage in insulated cables, physical stress or other causes.
The final effect is that the electrical properties of the damaged
cable section change, resulting in different values of $\mathbf{Y_{C}}$
and $\mathbf{\Gamma}$. In such a case, $\mathbf{A}$ is not simply
an extra transfer function added to the system, but a modified version
of the $\mathbf{H_{n}}$ relative to the damaged section. Hence, supposing
that the cable is uniformly damaged, a distributed fault causes two
new discontinuities, one at the beginning and one at the end of the
damaged section, which in turn results in a new series of peaks in
the time domain response. 

\begin{figure}[tb]
\begin{centering}
\subfloat[Frequency domain]{\begin{centering}
\includegraphics[width=1\columnwidth]{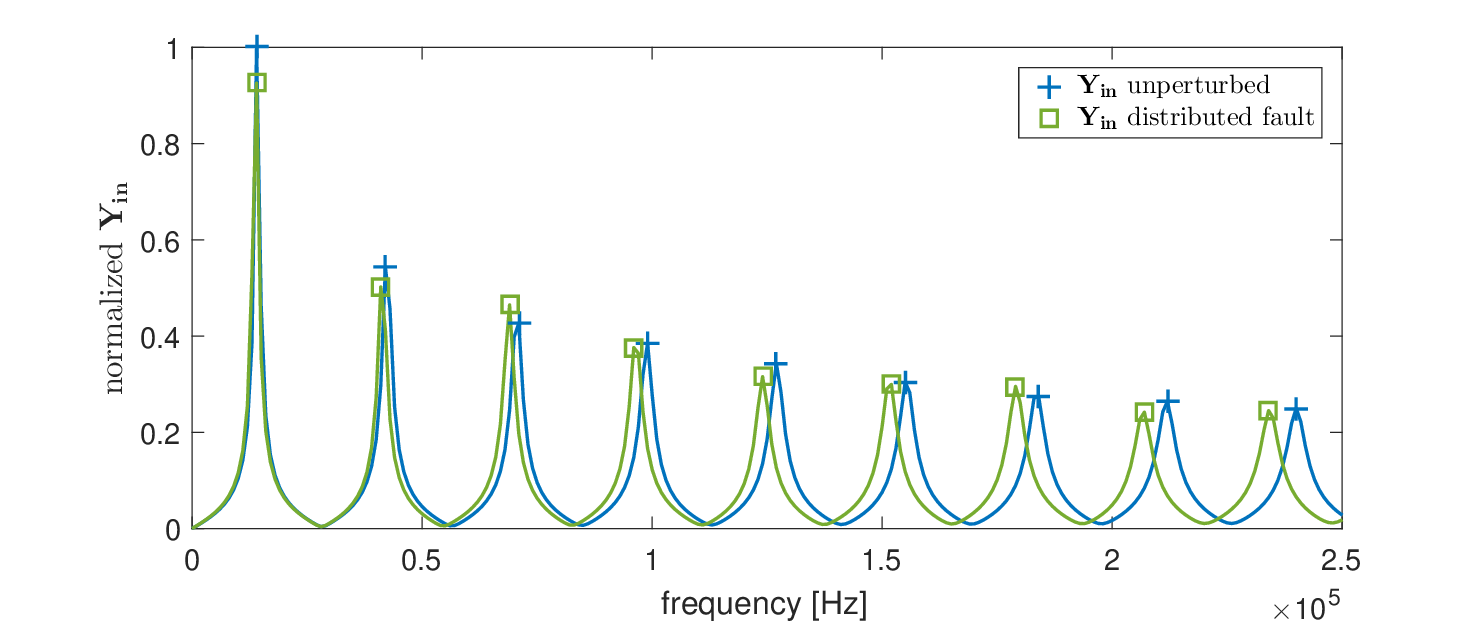}
\par\end{centering}

}
\par\end{centering}

\centering{}\subfloat[Time domain]{\begin{centering}
\includegraphics[width=1\columnwidth]{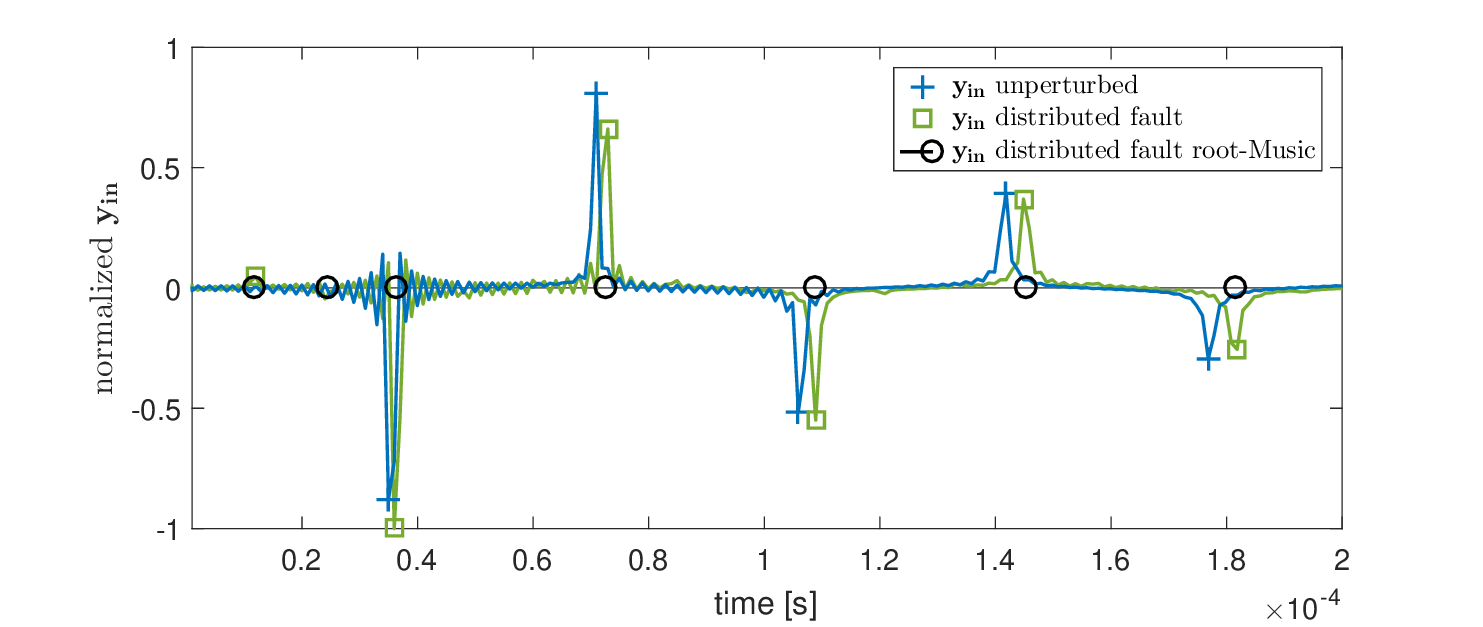}
\par\end{centering}

}\caption{$\mathbf{Y_{in}}$ simulated in presence of a distributed fault in
frequency (a) and time (b) domain.\label{fig:-simulated-in-1}}
\end{figure}

Considering again \eqref{eq:Htotal} and assuming the block $\mathbf{H_{j,0}}$
to be uniformly damaged, $\mathbf{Y_{C_{j,0}}}$, $\bm{\rho}_{\mathbf{L_{j,0}}}$
and $\mathbf{\Gamma_{j,0}}$ change w.r.t. the unperturbed condition.
The first two parameters cause, as concentrated faults, two new discontinuities:
one at the beginning and one at the end of the damaged section, which
in turn result in a new series of peaks in the time domain response.
Like for the concentrated fault case, this causes in frequency domain
the presence of an additional oscillating mode, its effect being more
evident at low frequencies. On the other hand, since variations of
$\mathbf{\Gamma_{j,0}}$ act on a series of exponential functions
which depend also on $\mathbf{\ell_{j,0}}$, they modify the previously
existing propagation modes of the signal. This results both in frequency
and in time domain in a shift of the local peaks, particularly at
high frequencies. The effect of a distributed fault on a single transmission
line is shown in Figure \ref{fig:-simulated-in-1}. 

Finally, distributed faults can be distinguished from concentrated
faults from their effect on the channel characteristics. Namely, distributed
faults introduce new propagation modes as concentrated faults do,
but also modify the existing ones. This difference is taken into account
in the algorithms presented in \cite{SGSII}. We also remark that,
while the effect of concentrated faults is more evident at low frequencies,
that of distributed faults is more evident at high frequencies. Therefore,
either different frequency ranges or highly broadband signals should
be used to sense the presence of both of them.

\subsection{Influence of the position of the anomaly}

The values of $\Delta_{ch}$ and $\Delta_{sup}$ depend on many factors,
among which the severity of the anomaly, its position inside the network,
the load values, the characteristics of the cables. It is not forcefully
said that a more severe anomaly or an anomaly located closer to the
receiver would produce higher values of $\Delta_{ch}$. Its values
depends on other factors too. 
\begin{figure}[tb]
\begin{centering}
\includegraphics[width=1\columnwidth]{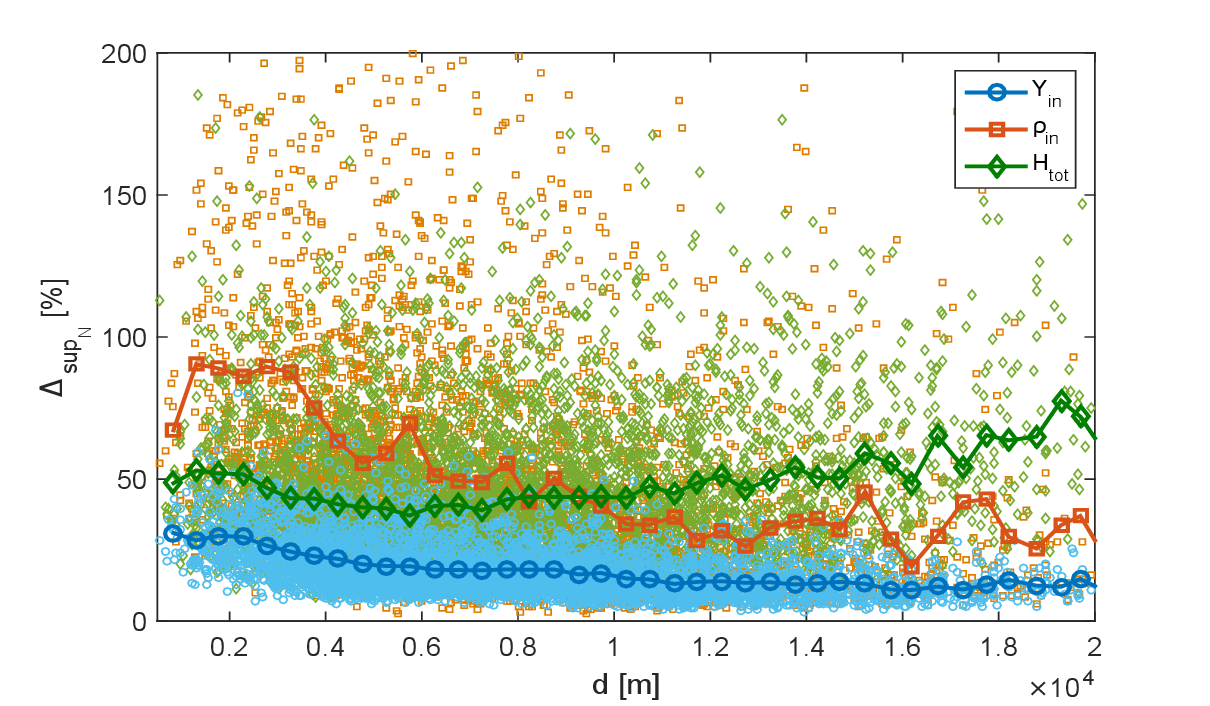}
\par\end{centering}

\caption{Effect of the distance $d$ of the anomaly from the receiver on $\Delta_{sup_{N}}$.\label{fig:Effect-of-the}}

\end{figure}
 However, we can state that, statistically, high values of $\Delta_{ch}$
and $\Delta_{sup}$ are expected for severe anomalies or high values
of the load impedances. The effect of the location of the anomaly
deserves a deeper analysis, that has been carried out by computing
$\Delta_{sup_{N}}$ for 10000 simulated networks affected by a randomly
located fault. For the details about the simulator please refer to
\cite{SGSII}. No noise has been considered in this case.

Fig. \ref{fig:Effect-of-the} shows that both $\Delta_{sup_{N}}^{rho}$
and $\Delta_{sup_{N}}^{Y}$ have an average declining trend with the
distance $d$ of the fault from the transmitter. This comes from the
fact that the amplitude of an echo signal naturally decreases with
the traveled distance. We also notice that, while the realizations
of $\Delta_{sup_{N}}^{Y}$ are rather close to the average, those
of $\Delta_{sup_{N}}^{rho}$ are more scattered. This is due to the
fact that often the values of $\bm{\rho}_{\mathbf{in}}$ are close
to 0 (see explanation in Section \ref{sub:General-models-of}).

More notably, Fig. \ref{fig:Effect-of-the} shows that the values
of $\Delta_{sup_{N}}^{H}$, after an initial decrease, tend to increase
again. We point out that we set the transmitter of the end-to-end
signal to be as far away as possible from the receiver. This means
that low values of $d$ correspond to faults located next to the receiver,
while high values of $d$ correspond to faults located next to the
transmitter. This means that a fault or, more in general, any anomaly
located either close to the transmitter or to the receiver, yields
higher values of $\Delta_{sup_{N}}^{H}$ compared to an anomaly located
in the middle of the network. To explain this fact, we rely again
on the symmetry of the end-to-end signals (see Theorem \ref{thm:TDS}).
Since the signal is unidirectional, an anomaly located close to the
transmitter would significantly modify it, but the result would be
smoothed out by the rest of the propagation path. Conversely, an anomaly
located close to the receiver acts on an already damped signal, but
the results gets to the receiver without further smoothing. Anomalies
located in the middle of the network suffer from both a weak incident
signal and a smoothed resulting signal delivered to the receiver.

\section{Conclusions\label{sec:Conclusions}}

In this paper, we presented a thorough analysis of what kind of information
can be retrieved about a power line network by using high-frequency
signals, such as those generated by power line modems. The proposed
approach enables to retrieve information about the network topology,
the aging of the cables, the variation of a load, the presence of
a fault. We provided closed-form formulas for the input impedance
and reflection coefficient at one node as well as for the CTF. We
then presented two different models, namely chain and superposition,
to represent the occurrence of a generic anomaly in the network. We
showed how these models can be applied to both reflectometric and
end-to-end sensing, and discussed the different results that they
can provide. We finally presented how different kind of anomalies,
namely a lumped fault, an impedance variation and a distributed fault,
affect the propagation of signals and discussed how it is possible
to distinguish between them using remote single-ended or double-ended
sensing. The interested reader may also read \cite{SGSII}, where
a complete measurement set-up to be included in PLC modems and different
algorithms are presented to automatically detect, classify and locate
electrical anomalies.

\bibliographystyle{IEEEtran}
\bibliography{femtocell_biblio}

 \newcommand{\noop}[1]{}
\begin{thebibliography}{10}
\providecommand{\url}[1]{#1}
\csname url@samestyle\endcsname
\providecommand{\newblock}{\relax}
\providecommand{\bibinfo}[2]{#2}
\providecommand{\BIBentrySTDinterwordspacing}{\spaceskip=0pt\relax}
\providecommand{\BIBentryALTinterwordstretchfactor}{4}
\providecommand{\BIBentryALTinterwordspacing}{\spaceskip=\fontdimen2\font plus
\BIBentryALTinterwordstretchfactor\fontdimen3\font minus
  \fontdimen4\font\relax}
\providecommand{\BIBforeignlanguage}[2]{{%
\expandafter\ifx\csname l@#1\endcsname\relax
\typeout{** WARNING: IEEEtran.bst: No hyphenation pattern has been}%
\typeout{** loaded for the language `#1'. Using the pattern for}%
\typeout{** the default language instead.}%
\else
\language=\csname l@#1\endcsname
\fi
#2}}
\providecommand{\BIBdecl}{\relax}
\BIBdecl

\bibitem{7961200}
A.~von Meier, E.~Stewart, A.~McEachern, M.~Andersen, and L.~Mehrmanesh,
  ``Precision micro-synchrophasors for distribution systems: A summary of
  applications,'' \emph{IEEE Transactions on Smart Grid}, vol.~8, no.~6, pp.
  2926--2936, Nov 2017.

\bibitem{5768099}
S.~Galli, A.~Scaglione, and Z.~Wang, ``For the grid and through the grid: The
  role of power line communications in the smart grid,'' \emph{Proceedings of
  the IEEE}, vol.~99, no.~6, pp. 998--1027, June 2011.

\bibitem{7467440}
C.~Cano, A.~Pittolo, D.~Malone, L.~Lampe, A.~M. Tonello, and A.~G. Dabak,
  ``State of the art in power line communications: From the applications to the
  medium,'' \emph{IEEE Journal on Selected Areas in Communications}, vol.~34,
  no.~7, pp. 1935--1952, July 2016.

\bibitem{1007375}
S.~Galli and D.~L. Waring, ``Loop makeup identification via single ended
  testing: beyond mere loop qualification,'' \emph{IEEE Journal on Selected
  Areas in Communications}, vol.~20, no.~5, pp. 923--935, Jun 2002.

\bibitem{neusphd}
C.~Neus, ``Reflectometric analysis of transmission line networks,'' Ph.D.
  dissertation, Vrije Universiteit Brussel, 2011.

\bibitem{333334d5-7017-4468-bab7-59871b51312a}
F.~Lindqvist, ``\BIBforeignlanguage{eng}{Estimation and detection of
  transmission line characteristics in the copper access network},'' Ph.D.
  dissertation, Lund University, 2011.

\bibitem{faultreview}
M.~Sedighizadeh, A.~Rezazadeh, and I.~Elkalashy, ``Approaches in high impedance
  fault detection -- a chronological review,'' \emph{Advances in Electrical and
  Computer Engineering}, vol.~10, no.~3, pp. 114--128, 2010.

\bibitem{7468545}
K.~Jia, T.~Bi, Z.~Ren, D.~W.~P. Thomas, and M.~Sumner, ``High frequency
  impedance based fault location in distribution system with dgs,'' \emph{IEEE
  Transactions on Smart Grid}, vol.~9, no.~2, pp. 807--816, March 2018.

\bibitem{sgc2017}
F.~Passerini and A.~M. Tonello, ``Full duplex power line communication modems
  for network sensing,'' in \emph{2017 IEEE International Conference on Smart
  Grid Communications (Smartgridcomm)}, October 2017.

\bibitem{6295693}
A.~N. Milioudis, G.~T. Andreou, and D.~P. Labridis, ``Enhanced protection
  scheme for smart grids using power line communications techniques -- part
  {II}: Location of high impedance fault position,'' \emph{IEEE Transactions on
  Smart Grid}, vol.~3, no.~4, pp. 1631--1640, Dec 2012.

\bibitem{6497543}
A.~M. Pasdar, Y.~Sozer, and I.~Husain, ``Detecting and locating faulty nodes in
  smart grids based on high frequency signal injection,'' \emph{IEEE
  Transactions on Smart Grid}, vol.~4, no.~2, pp. 1067--1075, June 2013.

\bibitem{Paul:2007:AMT:1554645}
C.~R. Paul, \emph{Analysis of Multiconductor Transmission Lines}, 2nd~ed.\hskip
  1em plus 0.5em minus 0.4em\relax Wiley-IEEE Press, 2007.

\bibitem{SGSII}
F.~Passerini and A.~M. Tonello, ``Smart grid network sensing using power line
  modems: A framework for anomaly detection and localization,'' \emph{Submitted
  to IEEE Transactions on Smat Grids}, 2018.

\bibitem{versolatto2011an}
F.~Versolatto and A.~Tonello, ``An {MTL} theory approach for the simulation of
  {MIMO} power-line communication channels,'' \emph{IEEE Transactions on Power
  Delivery}, vol.~26, no.~3, pp. 1710--1717, 2011.

\bibitem{book:538623}
P.~D.~B. Ronald E.~Miller, \emph{Input-Output Analysis: Foundations and
  Extensions - 2nd edition}, 2nd~ed.\hskip 1em plus 0.5em minus 0.4em\relax
  Cambridge University Press, 2009.

\bibitem{6312388}
P.~I. Somlo and J.~D. Hunter, ``Condition for reflection coefficient magnitude
  greater than one for passive transmission line and passive load,'' \emph{IEEE
  Transactions on Instrumentation and Measurement}, vol. IM-30, no.~3, pp.
  230--231, Sept 1981.

\bibitem{tonello2012a}
A.~M. Tonello, F.~Versolatto, B.~B{\'e}jar, and S.~Zazo, ``A fitting algorithm
  for random modeling the plc channel,'' \emph{IEEE Transactions on Power
  Delivery}, vol.~27, no.~3, pp. 1477--1484, Jul. 2012.

\bibitem{stoica2005spectral}
\BIBentryALTinterwordspacing
P.~Stoica and R.~Moses, \emph{Spectral Analysis of Signals}.\hskip 1em plus
  0.5em minus 0.4em\relax Pearson Prentice Hall, 2005. [Online]. Available:
  \url{https://books.google.at/books?id=h78ZAQAAIAAJ}
\BIBentrySTDinterwordspacing

\bibitem{ahmed2012topology2}
M.~Ahmed and L.~Lampe, ``Parametric and nonparametric methods for power line
  network topology inference,'' in \emph{Power Line Communications and Its
  Applications (ISPLC), 2012 16th IEEE International Symposium on}, March 2012,
  pp. 274--279.

\bibitem{tonello2011bottomup}
A.~Tonello and F.~Versolatto, ``Bottom-up statistical {PLC} channel modelling -
  part \uppercase{I}: Random topology model and efficient transfer function
  computation,'' \emph{IEEE Transactions on Power Delivery}, vol.~26, no.~2,
  pp. 891--898, Apr. 2011.

\bibitem{7878929}
A.~Codino, Z.~Wang, R.~Razzaghi, M.~Paolone, and F.~Rachidi, ``An alternative
  method for locating faults in transmission line networks based on time
  reversal,'' \emph{IEEE Transactions on Electromagnetic Compatibility},
  vol.~59, no.~5, pp. 1601--1612, Oct 2017.

\bibitem{4200703}
S.~Kay, ``Optimal signal design for detection of gaussian point targets in
  stationary gaussian clutter/reverberation,'' \emph{IEEE Journal of Selected
  Topics in Signal Processing}, vol.~1, no.~1, pp. 31--41, June 2007.

\bibitem{soltanalian}
M.~Soltanalian, ``\BIBforeignlanguage{eng}{Signal design for active sensing and
  communications},'' Ph.D. dissertation, Uppsala University, 2014.

\bibitem{abboud}
L.~Abboud, ``\BIBforeignlanguage{eng}{Time reversal techniques applied to wire
  fault detection and location in wire networks},'' Ph.D. dissertation,
  {SUPELEC} - Univ Paris-Sud, 2013.

\bibitem{amhifatl}
S.~Maximov, V.~Torres, H.~F. Ruiz, and J.~L. Guardado, ``Analytical model for
  high impedance fault analysis in transmission lines,'' \emph{Mathematical
  Problems in Engineering}, vol. 2014, 2014.

\bibitem{6311450}
S.~Gautam and S.~M. Brahma, ``Detection of high impedance fault in power
  distribution systems using mathematical morphology,'' \emph{IEEE Transactions
  on Power Systems}, vol.~28, no.~2, pp. 1226--1234, May 2013.

\bibitem{880817}
S.~Galli, ``Exact conditions for the symmetry of a loop,'' \emph{IEEE
  Communications Letters}, vol.~4, no.~10, pp. 307--309, Oct 2000.

\end{thebibliography}

\clearpage

\appendices{}

\section{Derivation of the input impedance in the general case \label{sec:Derivation-of-the}}

In this Appendix, we show how \eqref{eq:Yingeneral} and \eqref{eq:pingeneral}
are derived starting from \eqref{eq:yinsingleexpanded} and \eqref{eq:pinsingleexpanded}.

\begin{figure}[tb]
\centering{}\includegraphics[width=0.9\columnwidth]{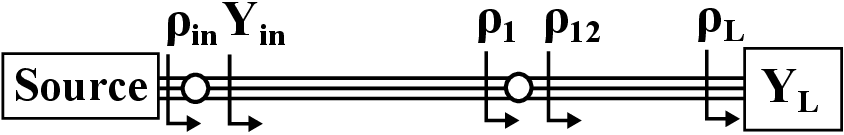}\caption{Example of a network with two sections and one load.\label{fig:Example-of-a}}
\end{figure}

In Section \ref{sec:Propagation-of-signals}, we derived the Taylor
series of $\mathbf{Y}_{\mathbf{in}}$ and $\bm{\rho}_{\mathbf{in}}$
in the case of a single line with a load at the end. Here we first
consider a MTL composed by two line sections of length $\ell_{1}$
and $\ell_{2}$ with no load at the junction and a load at the end
(see Fig. \ref{fig:Example-of-a}), derive $\mathbf{Y}_{\mathbf{in}}$
and $\bm{\rho}_{\mathbf{in}}$ and finally derive by induction.

The problem of finding $\mathbf{Y}_{\mathbf{in}}$ and $\bm{\rho}_{\mathbf{in}}$
for the network depicted in Fig. \ref{fig:Example-of-a} reduces to
the problem of finding $\bm{\rho}_{\mathbf{1}}$, which is the equivalent
load reflection coefficient relative to the first line section. Then
\eqref{eq:yinsingleexpanded} and \eqref{eq:pinsingleexpanded} can
be applied using $\bm{\rho}_{\mathbf{1}}$ instead of \eqref{eq:plsingle}.
$\bm{\rho}_{\mathbf{1}}$ is also the input reflection coefficient
w.r.t. the second line, therefore we can write
\begin{multline}
\bm{\rho}_{\mathbf{1}}=\mathbf{N_{2}}\mathbf{T_{2}}\left[\bm{\rho}_{\mathbf{12}}^{\mathbf{M}}+\left(\mathbf{I}-\bm{\rho}_{\mathbf{G}}^{\mathbf{M}}\bm{\rho}_{\mathbf{G}}^{\mathbf{M}}\right)\bm{\rho}_{\mathbf{B}\mathbf{2}}^{\mathbf{M}}\right]\mathbf{T_{2}}^{-1}\mathbf{N_{2}}^{-1}+\\
\mathbf{N_{2}}\mathbf{T_{2}}\Bigg[\left(\mathbf{I}-\bm{\rho}_{\mathbf{12}}^{\mathbf{M}}\bm{\rho}_{\mathbf{12}}^{\mathbf{M}}\right)\bm{\rho}_{\mathbf{B}\mathbf{2}}^{\mathbf{M}}\sum_{n=1}^{\infty}\left(-1\right)^{n}\left(\bm{\rho}_{\mathbf{B}\mathbf{2}}^{\mathbf{M}}\right)^{n}\Bigg]\mathbf{T_{2}}^{-1}\mathbf{N_{2}}^{-1}\label{eq:p1}
\end{multline}
where $\bm{\rho}_{\mathbf{12}}^{\mathbf{M}}=\mathbf{T}^{-1}\mathbf{Y}_{\mathbf{C_{2}}}\left(\mathbf{Y}_{\mathbf{C_{2}}}+\mathbf{Y}_{\mathbf{C_{1}}}\right)^{-1}\left(\mathbf{Y}_{\mathbf{C_{2}}}-\mathbf{Y}_{\mathbf{C_{1}}}\right)\mathbf{Y}_{\mathbf{C_{2}}}^{-1}\mathbf{T}$,
$\mathbf{N_{2}}=\left(\mathbf{Y}_{\mathbf{C_{1}}}+\mathbf{Y}_{\mathbf{C_{2}}}\right)\mathbf{Y}_{\mathbf{C_{2}}}^{-1}$
and $\bm{\rho}_{\mathbf{B}\mathbf{2}}^{\mathbf{M}}=\mathbf{e}^{-\mathbf{\Gamma_{2}}\ell_{2}}\bm{\rho}_{\mathbf{L}}^{\mathbf{M}}\mathbf{e}^{-\mathbf{\Gamma_{2}}\ell_{2}}$.
We now substitute $\bm{\rho}_{\mathbf{L}}^{\mathbf{M}}$ in \eqref{eq:yinsingleexpanded}
and \eqref{eq:pinsingleexpanded} with $\bm{\rho}_{\mathbf{1}}^{\mathbf{M}}$,
thus obtaining
\begin{multline}
\mathbf{Y}_{\mathbf{in}}=\mathbf{T_{1}}\Bigg[\mathbf{I}+2\mathbf{e}^{-\mathbf{\mathbf{\Gamma}_{1}}\ell_{1}}\mathbf{O_{2}}\bm{\rho}_{\mathbf{12}}^{\mathbf{M}}\mathbf{O_{2}}^{-1}\mathbf{e}^{-\mathbf{\mathbf{\Gamma}_{1}}\ell_{1}}+\\
2\mathbf{e}^{-\mathbf{\mathbf{\Gamma}_{1}}\ell_{1}}\mathbf{O_{2}}\mathbf{e}^{-\mathbf{\Gamma_{2}}\ell_{2}}\bm{\rho}_{\mathbf{L}}^{\mathbf{M}}\mathbf{e}^{-\mathbf{\Gamma_{2}}\ell_{2}}\mathbf{O_{2}}^{-1}\mathbf{e}^{-\mathbf{\mathbf{\Gamma}_{1}}\ell_{1}}+\\
2\sum_{n=2}^{\infty}\dots\Bigg]\mathbf{T_{1}}^{-1}\mathbf{Y}_{\mathbf{C}}\label{eq:yin2}
\end{multline}
and 
\begin{multline}
\bm{\rho}_{\mathbf{in}}=\mathbf{N_{1}}\mathbf{T_{1}}\Bigg[\bm{\rho}_{\mathbf{G_{1}}}^{\mathbf{M}}+\mathbf{Q_{1}}\Big(\mathbf{e}^{-\mathbf{\mathbf{\Gamma_{1}}}\ell_{1}}\mathbf{O_{2}}\bm{\rho}_{\mathbf{12}}^{\mathbf{M}}\mathbf{O_{2}}^{-1}\mathbf{e}^{-\mathbf{\mathbf{\Gamma_{1}}}\ell_{1}}+\\
\mathbf{e}^{-\mathbf{\mathbf{\Gamma}_{1}}\ell_{1}}\mathbf{O_{2}}\mathbf{Q_{2}}\mathbf{e}^{-\mathbf{\Gamma_{2}}\ell_{2}}\bm{\rho}_{\mathbf{L}}^{\mathbf{M}}\mathbf{e}^{-\mathbf{\Gamma_{2}}\ell_{2}}\mathbf{O_{2}}^{-1}\mathbf{e}^{-\mathbf{\mathbf{\Gamma}_{1}}\ell_{1}}\Big)+\\
\sum_{n=1}^{\infty}\dots\Bigg]\mathbf{T_{1}}^{-1}\mathbf{N_{1}}^{-1}\label{eq:pin2}
\end{multline}
where $\mathbf{O_{2}}=\mathbf{T_{2}}^{-1}\mathbf{N_{2}}\mathbf{T_{2}}$
and $\mathbf{Q_{n}}=\mathbf{I}-\bm{\rho}_{\mathbf{G_{n}}}^{\mathbf{M}}\bm{\rho}_{\mathbf{G_{n}}}^{\mathbf{M}}$.
Similarly to the single line case, \eqref{eq:yin2} and \eqref{eq:pin2}
are composed by different terms. The first is a constant w.r.t. the
line lengths that depends on the impedance mismatch between the source
and the first line segment. Then two terms follow, which depend exponentially
on $2\ell_{1}$ and on $2\left(\ell_{1}+\ell_{2}\right)$ respectively.
Finally, an infinite series follows (not explicitly written here)
with terms that depend on multiples of $2\ell_{1}$, $2\left(\ell_{1}+\ell_{2}\right)$
and combinations of these two terms. 

The same reasoning can be extended by induction to lines composed
by any number of segments, also with multiple segments branched to
a single node, thus yielding the results expressed by \eqref{eq:Yingeneral}
and \eqref{eq:pingeneral}.

\section{Proof of Theorem \ref{thm:TDS} \label{sec:Proof-of-Theorem}}

We explained in Section \ref{sub:End-to-end} that the transfer function
by any two points in a wired network is given by \eqref{eq:Htotal}.
We define now the backbone as the shortest path between the transmitter
and the receiver nodes, that is either the line of sight path in wireless
networks or the shortest sequence of line segments connecting the
two ends in wired networks. We also define $\mathbf{H_{tot}^{A\rightarrow B}}$
and $\mathbf{H_{tot}^{B\rightarrow A}}$ as the transfer functions
from A to B and from B to A respectively.

$\mathbf{H_{tot}^{A\rightarrow B}}$can be computed using the carry-back
method described in \cite{versolatto2011an}, which involves as first
step carrying back all the loads at the termination nodes to the backbone.
Since every signal from A to B or from B to A travels through the
same backbone, this first step yields the same result also for $\mathbf{H_{tot}^{B\rightarrow A}}$
. At this point the network has been reduced to a sequence of line
segments with equivalent loads branched at every node. In the following
step, the resulting $\mathbf{H_{n}}$ matrices are multiplied with
each other in order to give the total transfer function. It has already
been demonstrated in \cite{880817} that if $\mathbf{Y_{R}^{A}}=\mathbf{Y_{R}^{B}}=\mathbf{Y_{L}^{A}}=\mathbf{Y_{L}^{B}}$,
then $\mathbf{H_{tot}^{A\rightarrow B}}=\mathbf{H_{tot}^{B\rightarrow A}}$,
otherwise the equality does not hold and in general the two transfer
functions are not similar to each other. However, \eqref{eq:Htotal}
shows that, apart from a multiplicative factor, the same exponentials
are multiplied with each other in the case of $\mathbf{H_{tot}^{A\rightarrow B}}$
and $\mathbf{H_{tot}^{B\rightarrow A}}$, just in opposite order.
When transforming \eqref{eq:Htotal} to time domain, this results
in a sequence of smoothed peaks whose location is the same.
\end{document}